\documentclass[aps,prb,twocolumn,showpacs,superscriptaddress]{revtex4}
\usepackage{bbm}
\usepackage{mathrsfs}
\usepackage{epsfig}
\usepackage{graphicx}
\usepackage{amsfonts}
\usepackage[figuresright]{rotating}
\usepackage{amssymb}
\usepackage{amsmath}
\usepackage{dcolumn}
\usepackage{bm}

\def\avg#1{\langle#1\rangle}

\def\be{\begin{equation}} \def\ee{\end{equation}}
\def\bea{\begin{eqnarray}} \def\eea{\end{eqnarray}}

\def\nn{\nonumber}

\begin{document}
\title{Quantum Monte Carlo simulation of thermodynamic properties of
SU($2N$) ultracold fermions in
optical lattices}
\author{Zhichao Zhou}
\affiliation{School of Physics and Technology, Wuhan University, Wuhan 430072, China}
\author{Zi Cai}
\affiliation{Institute for Quantum Optics and Quantum Information,
 Austrian Academy of Sciences, 6020 Innsbruck, Austria}
\author{Congjun Wu}
\affiliation{Department of Physics, University of California, San Diego, CA 92093, USA}
\author{Yu Wang}
\email{yu.wang@whu.edu.cn}
\affiliation{School of Physics and Technology, Wuhan University, Wuhan 430072, China}

\begin{abstract}
We have systematically studied the thermodynamic properties of
a two-dimensional half-filled SU($2N$) Hubbard model on a square lattice
by using the determinant quantum Monte Carlo method.
The entropy-temperature relation, the isoentropy curve, and
the probability distribution of the onsite occupation number are calculated
in both SU(4) and SU(6) cases, which exhibit prominent features
of the Pomeranchuk effect.
We analyze these thermodynamic behaviors based on charge and spin energy scales.
In the charge channel, the interaction strength that marks the crossover from the
weak to strong interaction regimes increases with the number of fermion components.
In the spin channel, increasing the number of fermion components
enhances quantum spin fluctuations, which is shown in the simulations of uniform spin susceptibilities
and antiferromagnetic structure factors.
\end{abstract}
\pacs{71.10.Fd, 03.75.Ss, 37.10.Jk,71.27.+a}
\maketitle


\section{Introduction}
In condensed matter physics, the goal of generalizing SU(2)
lattice fermion or spin models to those with high symmetries of
SU($N$) \cite{Affleck1988, Arovas1988,Read1990} or Sp($N$)
\cite{Read1991,Sachdev1991}, was originally to employ the
systematic $1/N$-expansion to handle strong correlation physics,
especially in the cases with doping or frustrations.
Generally speaking, the large symmetries of SU($N$) and Sp($N$)
enhance quantum spin fluctuations and suppress the antiferromagnetic
(AF) order
\cite{Arovas1988,Affleck1988,Sachdev1991}.
Various exotic quantum paramagnetic phases have been proposed based on the
large-$N$ method, including various valence bond solid states and
quantum spin liquid states \cite{Harada2003,Assaad2005,Corboz2011,
Corboz2012b,Paramekanti2007}. However, in conventional solid states,
the SU($N$) symmetry is rare and thus the SU($N$) Hubbard or Heisenberg models
are purely of academic interest.

With the rapid development of the ultracold atom experiments, the realization of
multi-component fermionic Hubbard models with the SU($2N$) or Sp($2N$)
symmetry has become a realistic goal (the number of fermion
components due to the hyperfine spin degree of freedom is
naturally an even number).
It was proposed that the simplest Sp($2N$) and SU($2N$) Hubbard models
with $2N=4$ can be realized in the hyperfine spin-$\frac{3}{2}$
alkali and alkaline-earth atoms \cite{Wu2003,Wu2006a}. In these spin-$\frac{3}{2}$ Hubbard models, an exact
Sp(4) spin symmetry exists without any fine-tuning of parameters, which is further enlarged to SU(4) when the interactions do not rely on hyperfine spin.
The alkaline-earth atoms, {\it e.g.}, $^{173}$Yb and $^{87}$Sr,
have a closed shell of valence electrons and thus their hyperfine spins are simply  nuclear spins.
The interactions between the atoms with different hyperfine spins are
insensitive to the nuclear spins, leading to the SU($2N$)
symmetry with $2N$ being the number of fermion components
\cite{Gorshkov2010,Wu2012}.

Recently, significant progress has been made in the experiment of ultracold alkaline-earth
fermions with large hyperfine spins. The $^{173}$Yb and $^{87}$Sr
atoms have been cooled down to quantum degenerate temperatures
\cite{Taie2010,Desalvo2010,Hideaki2011}, revealing the
SU(6) and SU(10) symmetries respectively.
Furthermore, an SU(6) single-band fermionic Hubbard model has also been realized
with $^{173}$Yb atoms in a three-dimensional optical lattice\cite{Taie2012}.
Beyond single-band case, the spin-exchange interactions have recently been observed in the two-orbital
SU($6$) and SU($10$) fermion system respectively\cite{Scazza2014,Zhang2014}.
Also the number of spin components can be tuned experimentally
\cite{Pagano2014}.
Theoretically, the novel symmetries of the multi-component Hubbard
model can give rise to the novel superfluidity \cite{Lecheminant2005,
Honerkamp2004,Honerkamp2004a,He2006,Cherng2007,Rapp2007,Rapp2008}
and exotic quantum magnetism \cite{Wu2005a,Cazalilla2009,Manmana2011,
Hung2011,Szirmai2011a,Szirmai2011b,Hermele2009,Hermele2011,
Xu2010,Heinze2013}.

In ultracold atom experiments, achieving low enough temperature regime below the
spin superexchange scale has been considered a benchmark
for simulating strongly correlated quantum systems.
Despite numerous efforts by experimentalists, achieving this temperature
regime is still out of reach and remains one of the most challenging
problems in this field. So far, ultracold fermions
in optical lattices have been cooled down to temperature regime below the
hopping energy scale, $T\sim t$. One of the promising schemes for further cooling the system into spin
superexchange scale $T\sim J$ is known as interaction-induced
adiabatic cooling\cite{Werner2005}, a cooling scheme by adiabatically increasing interactions.
This cooling scheme utilizes the Pomeranchuk effect which was originally proposed in $^3$He systems.
However, for a two-component Hubbard model in conventional lattices,
the Pomeranchuk effect is weak due to the antiferromagnetic correlations in the
SU(2) Mott insulator.
It is still controversial whether the system
can be cooled down to spin superexchange temperature by Pomeranchuk
cooling\cite{Werner2005,Dare2007,Paiva2010,Paiva2011}.
As we show below, the multi-component
SU($2N$) Hubbard model significantly facilitates the Pomeranchuk cooling,
cooling the system down to the temperature scale of $J$ from an
initial temperature that is currently accessible in experiments.

This paper extends the previous work reported in
Ref. [\onlinecite{Cai2013}].
We have performed detailed determinant quantum Monte Carlo (DQMC)
simulations of thermodynamic properties of the half-filled
SU($2N$) Hubbard model with $2N=4$ and 6 in the temperature
regime $J<T<t$.
We calculated the entropy-temperature relation and isoentropy curves, which show the enhancement of entropy with
increasing interaction strength in the intermediate temperature regime,
{\it i.e.}, the Pomeranchuk effect.
The probability distributions of the onsite occupation number
show the enhancement of particle localization as
temperature increases in the low and intermediate temperature regimes.
The uniform spin susceptibilities and AF structure
factors are also calculated.

The rest part of this paper is organized as follows.
In Section \ref{sect:model}, we introduce the definition of the
SU($2N$) Hubbard model.
A discussion of the charge and spin energy scales of the
half-filled SU($2N$) Hubbard model is followed in
Section \ref{sect:energyscale}.
The parameters of the DQMC simulations are given in Section
\ref{sect:parameters}.
In Section \ref{sect:thermo}, we present the results of DQMC study on
the thermodynamic properties of the half-filled SU(4) and SU(6) Hubbard models.
In Section \ref{sect:mag}, the magnetic properties at
finite temperatures are investigated.
Conclusions are drawn in Section \ref{sect:conclusion}.

\section{The SU($2N$) Hubbard model}
\label{sect:model}

Since naturally the number of fermion components due to the hyperfine
spin degree of freedom is an even number, we only consider Hubbard
model with SU($2N$) symmetry.
At half-filling, an SU($2N$) Hubbard model
is defined by the lattice Hamiltonian:
\begin{equation}
H=-t\sum_{\avg{i,j},\alpha}\Big\{
c^\dag_{i\alpha}c_{j\alpha}+h.c.\Big\} +\frac{U}{2}\sum_i\big(n_i-N
\big)^2,
\label{eq:sun}
\end{equation}
where $\avg{i,j}$ denotes nearest neighbors and the sum runs over sites of a two-dimensional square lattice;
$\alpha$ represents spin indices running from 1 to $2N$; $n_i$ is the particle number operator on site $i$ defined by
$n_i=\sum_{\alpha=1}^{2N} c^\dag_{i\alpha}c_{i\alpha}$;
$t$ and $U$ are the nearest neighbor hopping integral and the onsite interaction, respectively.

This definition of Hubbard Hamiltonian Eq.(\ref{eq:sun}) offers several
advantages.
In the atomic limit ($t=0$), consider a half-filled lattice
with $N$ particles per site, the energy cost of moving a particle
from one site to its neighboring site is $U$,
which is independent of $N$.
Due to half-filling, the chemical potential $\mu$ vanishes in this grand canonical Hamiltonian.
Eq.(\ref{eq:sun}) also has particle-hole symmetry in bipartite
lattices, which removes the sign
problem in DQMC simulations for an arbitrary value of $2N$.

In terms of the multiplets of SU($2N$) fermions in the fundamental
representation, the generators of the SU($2N$) group can be written as
\begin{equation}
S_{\alpha\beta}(i)=c^\dag_{\alpha}(i)c_{\beta}(i)-
\frac{\delta^{\alpha\beta}} {2N}\sum_{\gamma=1}^{2N}
c^\dag_\gamma(i) c_\gamma(i), \label{eq:su2n_gen}
\end{equation}
where $\alpha$ and $\beta$ run from 1 to $2N$.
The generators defined above are not independent of each other, since
the diagonal operators satisfy the relation,
$\sum_\alpha S_{\alpha\alpha}(i)=0$.
Nevertheless, the definition of operators, Eq.(\ref{eq:su2n_gen}), results in a simple commutation relation
\begin{equation}
[S^{\alpha\beta}, S^{\gamma\delta}]=\delta_{\beta\gamma}
S^{\alpha\delta} -\delta_{\alpha\delta} S^{\gamma\beta}.
\end{equation}

For convenience, we define the structure factor $S_{su(2N)}(\vec q)$
as
\bea
S_{su(2N)}(\vec{q})=\frac{1}{L^2}\sum_{i,j} e^{i\vec{q}
\cdot \vec r} S_{spin}(i,j),
\label{eq:struc}
\eea
where $\vec r$ is the relative vector between sites $i$ and $j$.
$S_{spin}(i,j)$ is the SU($2N$) version of equal-time
spin-spin correlation functions defined by
\bea
S_{spin}(i,j)=\frac{1}{(2N)^2-1} \sum_{\alpha,\beta}\langle
S_{\alpha\beta,i} S_{\beta\alpha,j}\rangle.
\label{SS}
\eea

\section{The charge and spin energy scales}
\label{sect:energyscale}

Before going to the results of DQMC simulation, let us present a qualitative
understanding of the physics of the half-filled Hubbard model on
a square lattice.
The charge channel and spin channel are characterized by two
energy scales $\Delta_c$ and $\Delta_s$, respectively,
which will be discussed in both the weak and strong
interaction regimes below.

\subsection{The weak interaction regime and the atomic limit}
\label{sect:weakatom}

We first consider the weak interaction limit ($U\rightarrow 0$).
In this regime, the underlying Fermi surface plays an important
role, which possesses the diamond shape and thus exhibits perfect nesting.
The key physics then arises from the Fermi surface nesting:
the spin susceptibility in the non-interacting limit divergences
logarithmically, and thus an infinitesimal repulsive interaction
generates AF long-range-order.
In this case, the gapped quasi-particle excitations carry both
charge and spin quantum numbers.
The charge and spin energy scales are identical in this regime as
\cite{Fradkin1991}
\bea
\Delta_s/t=\Delta_c/t= 4\pi^2e^{-\sqrt{2\pi t/U}}.
\eea
Certainly, in the weak interaction regime, the system is a weak insulator.
Although it is gapped, charge fluctuations cannot be neglected.
On the contrary, in the atomic limit ($t=0$, or,
$U\rightarrow +\infty$), the Fermi surface completely disappears, and
we need to use the local moment picture.
At zero temperature, charge fluctuations are completely frozen.
In the Mott-insulating state, we use the single-particle gap
to denote the charge fluctuations, {\it i.e.}, the energy cost by adding or
removing a particle from the Mott-insulating state.
It is the energy difference between the two
energy levels $\Delta_c=E_{n_i=N+1}- E_{n_i=N}$
which equals $\frac{U}{2}$ in the atomic limit.
Since the hopping process is completely suppressed, the AF exchange
energy $J=0$, {\it i.e.}, the spin energy scale $\Delta_s$ is zero
in the atomic limit.

\subsection{The strong interaction regime}

\begin{figure}[htb]
\includegraphics[width=0.7\linewidth]{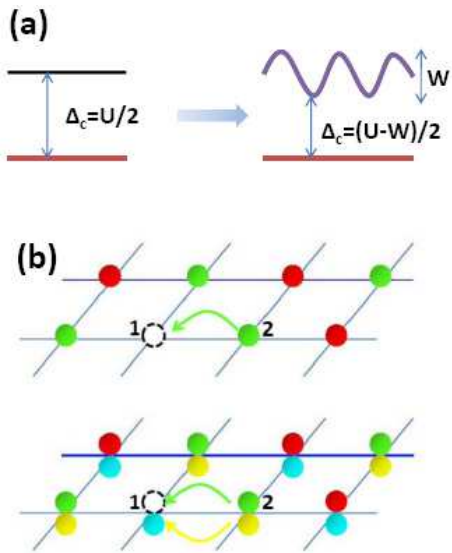}
\caption{(a) Energy dispersion of a charge excitation
on the background of the half-filled Mott-insulating state:
(left) the atomic limit with $t/U=0$, (right) $t/U\ll 1$.
(b) Sketches of a hole hopping in the SU(2) (up) and SU(4)
(down) AF backgrounds, respectively.
(Fig \ref{fig:hopping} (b) is from Ref. [\onlinecite{Cai2013}].
)
}
\label{fig:hopping}
\end{figure}

Let us consider the strong interaction regime in which $U$
is large but not strong enough to completely suppress charge fluctuations,
{\it i.e.}, $t/U\ll 1$.
Let us consider the charge energy scale $\Delta_c$ by adding an
extra particle (hole) onto the background of the Mott-insulating state.
As shown in Fig. \ref{fig:hopping} (a), the propagation of the particle
(hole) expands the excitation energy level at $U/2$ in the atomic
limit into an energy band.
The charge gap corresponds to the band bottom and thus is lowered to
\bea
\Delta_c=\frac {U}{2}-\frac {W}{2},
\eea
where  $W$ is the band width and is determined by the hopping process of
the extra particle (hole).

In Fig. \ref{fig:hopping} (b), we compare the hopping process of
an extra hole in the half-filled SU(2) and SU(4) Mott insulators.
In the SU(4) case, there are much more routes for the hole to hop
from one site to its neighboring site than it does in the SU(2) case.
Typically speaking, the number of hopping process for an extra
 particle or hole under the half-filled Mott-insulating background scales as $N$,
thus we estimate $W \propto Nt$.
The mobility of the extra hole is greatly enhanced in
the SU($2N$) Mott-insulating state.
Consequently, the charge energy scale $\Delta_c$ is significantly
lowered with increasing $2N$.
Naively, we could estimate that $\Delta_c$ vanishes at
\bea
U_c \approx Nt.
\label{eq:Uc}
\eea
Certainly, due to Fermi surface nesting, $\Delta_c$ does not
vanish even in the weak interaction regime but becomes exponentially
small.
Nevertheless, $U_c$ sets up a scale of interaction strength
to separate the regimes of the Fermi surface nesting and
the local moments.

The low energy physics in the strong interaction regime is described by
the SU($2N$) generalization of the Heisenberg model:
\begin{equation}
H=J\sum_{\alpha,\beta,[
\vec{i}\vec{j}]} S_{\alpha\beta,\vec{i}}S_{\beta\alpha,\vec{j}},
\end{equation}
where the SU($2N$) spin operators are defined in Eq.(\ref{eq:su2n_gen}).
$J$ describes the strength of spin superexchange energy scale
which can be viewed as the spin energy scale $\Delta_s$.
The second order perturbation theory yields $J=4t^2/U$
which decreases as $U$ increases.
Noting that $\Delta_s\approx 4\pi^2 t e^{-2\pi\sqrt{t/U}}$ in
the weak interaction regime, which increases with $U$,
there should exist a peak in the intermediate
interaction regime.

\subsection{Effect of finite temperatures}

When the fermion system is deep in the Mott-insulting state, in the low temperature regime
$T\ll\Delta_s$, charge fluctuations are strongly suppressed,
and the physics is dominated by spin superexchange process.
Therefore quantum spin fluctuations play an important role in determining the
magnetic properties of the Mott-insulating state at low temperatures.
On the contrary, at high temperatures $T\gg \Delta_c$, quantum
fluctuations give way to thermal fluctuations, which suppress
quantum correlations, and thus interaction effects can be
neglected.
In the intermediate temperature regime $\Delta_s < T < \Delta_c$,
$T$ is high enough to suppress the AF correlations, but not
sufficient to defreeze charge fluctuations.
Both quantum and thermal fluctuations are important in the intermediate temperature regime,
and the interplay between them gives rise to interesting phenomena
and universal properties \cite{Hazzard2013}.

\section{The DQMC method}
\label{sect:parameters}

The DQMC method is a widely used non-perturbative method for studying
strongly correlated fermion systems \cite{Blankenbecler1981,
Hirsch1983,Hirsch1985,White1989,Santos2003,Scalapino2006,
Binder1997,Suzuki1986}.
Provided that there is no sign problem, DQMC is known to be a
well-controlled and unbiased method, which yields asymptotically
exact results.
One of the most remarkable DQMC results is the AF
long-range order in the ground state of 2D SU(2) half-filled Fermi-Hubbard model
on a square lattice \cite{Hirsch1985,Hirsch1989,White1989}.
In the subsequent sections, we will use the DQMC method to simulate
thermodynamic properties of the half-filled SU($2N$) Hubbard model in
different regimes of temperature $T$ and interaction $U$.
We will also show how those results are related to the two energy
scales $\Delta_s$ and $\Delta_c$ analyzed in previous section.

Considering the error accumulation from matrix multiplications and
simulation time, the lowest temperature in simulations is set
as $T_L/t=0.1$ (with $\beta=t/T_L=10$).
The Suzuki-Trotter decomposition is used in which the error is proportional
to the cube of time discretization parameter $(\Delta\tau)^3$.
$\Delta\tau$ is set from $0.02$ to $0.05$ in the temperature regime
$T_L/t<T/t<0.5$, and the convergence with respect to the scalings of
$\Delta\tau$ has been checked.

A Hubbard-Stratonovich (HS) transformation that decomposes the onsite
Hubbard interaction term in the density channel \cite{Assaad2005}
preserves the SU($2$) symmetry of the Hubbard model, which can
also be generalized to the SU($2N$) Hubbard model.
The discrete HS decomposition with an Ising field only applies
to the spin-$\frac{1}{2}$ case \cite{Hirsch1983,Hirsch1985}.
For the cases of SU(4) and SU(6), an exact discrete decomposition
has been developed in Ref. [\onlinecite{Wang2014}], which is
explained in Appendix and employed in our simulations.
The simulated system is a $L\times L$ square lattice with $L=10$.
We focus on the parameter regimes of $0.1<T/t <10$ and $2\le U/t \le 12$.
For a typical data point, we use 10 QMC bins each of which includes 2000 warm-up steps
and 8000 measurement steps. We collect data once in each time slice.
In our simulations, $t$ is set to unity and
then the Hubbard $U$ and temperature $T$ are given in the unit of $t$.

\section{Thermodynamic properties of half-filled SU($2N$)
Hubbard model}
\label{sect:thermo}

In this section, we present the simulations of thermodynamic
properties of the half-filled SU($2N$) Hubbard model with $2N=4$ and $6$
on a square lattice, including the
entropy-temperature relation, and the Pomeranchuk effect.

\subsection{The entropy-temperature relation}
\label{sect:entropy}

\begin{figure}[htb]
\includegraphics[width=0.89\linewidth]{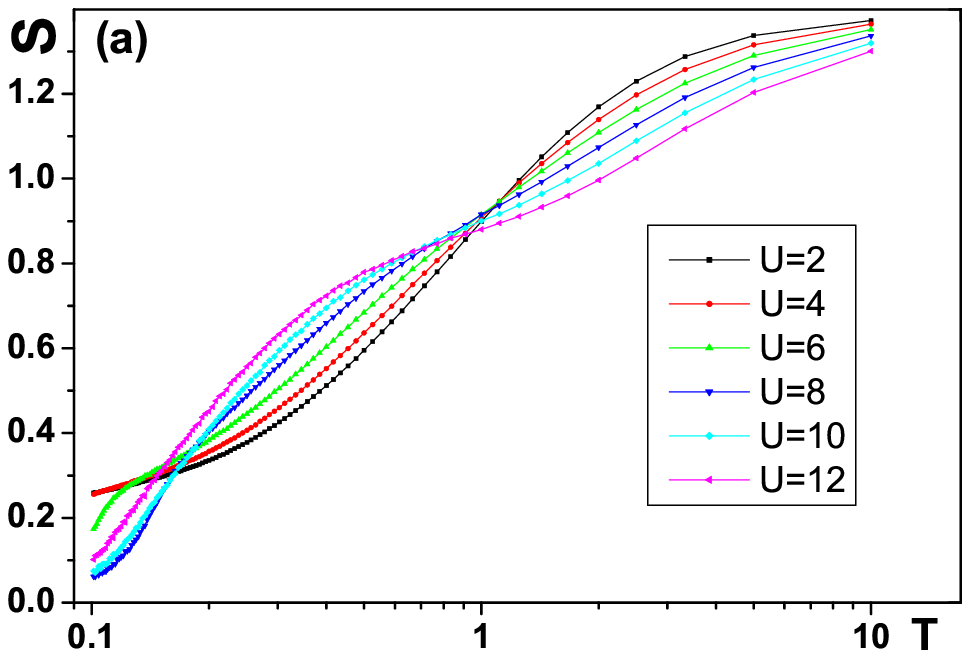}
\includegraphics[width=0.89\linewidth]{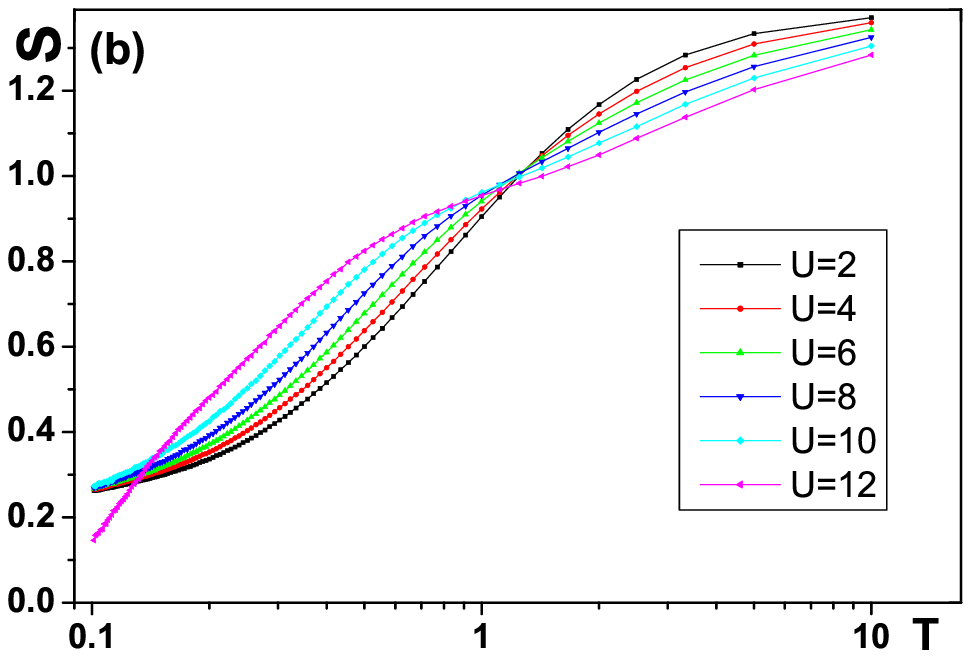}
\caption{The entropy per particle as a function of $T$ for
different values of $U$ in (a) SU(4) and (b) SU(6)
Hubbard models.
The system size is $L\times L$ with $L=10$.}
\label{fig:stu}
\end{figure}

In cold atom experiments, entropy is actually a
more physical quantity than temperature to characterize the system.
Below we present the simulated entropy in the half-filled
SU($4$) and SU($6$) Hubbard models.
The parameters of simulation are chosen in the regimes of $0.1 <T/t <10$
and $2\le U/t \le 12$, which are of interest in experiment.
The simulated entropy per particle (not per site) is defined by
$S_{SU(2N)}=S/(N L^2)$, where $S$ is the total entropy in the lattice.
The following formula is employed to calculate $S_{SU(2N)}$:
\bea
\frac{S_{SU(2N)}(T)}{k_B}=\ln 4+\frac{E(T)}T-
\int^\infty_T dT^\prime \frac{E(T^\prime)}{T^{\prime 2}},
\eea
where $\ln 4$ is the entropy at infinite temperature; $E(T)$
denotes the internal energy per particle at temperature $T$.

In Fig. \ref{fig:stu}, we show the entropy (per particle) of
SU(4) and SU(6) fermions as a function of $T$ for various
values of $U$.
In both cases, the curves cross at two {\it typical} temperatures $T_l$ (low)
and $T_h$ (high) which divide the temperature into three different regimes.
Let us first look at the SU(4) case as follows.

{\it The low temperature regime $T_L<T<T_l$}
In this regime, the dependence of the entropy per particle
$S_{SU(2N)}$ on $U$ is non-monotonic, which can be understood
by the competition between the spin energy scale $\Delta_s$
and the charge energy scale $\Delta_c$ as explained below.

For weak interactions $U/t<4$, $S_{SU(4)}$ is insensitive to $U$.
As explained in section \ref{sect:weakatom}, the physics
in this regime is characterized by the Fermi
surface nesting
$\Delta_c$ and $\Delta_s$ which are equal and are smaller than
$T_L/t\approx 0.1$ (the lowest temperature reached in our simulations), and thus
interaction effects are unimportant.
$S_{SU(4)}$ is then approximately the same as that in the
non-interacting limit.
The non-zero residue entropy is due
to the finite size effect which is caused by the degeneracy of single particle states right
located on the Fermi surface.
As $U$ increases, $\Delta_{c}$ increases faster than $\Delta_s$,
while $\Delta_s$ quickly reaches its maximum. In this interaction regime.
the relation $\Delta_c>\Delta_s>T$ holds,
and thus increasing $U$ freezes charge fluctuations but enhances
AF correlations.
Even though there cannot be true long-range AF ordering at finite
temperatures in 2D, the AF correlation length scales
as $e^{-\frac{\Delta_s}{T}}$.
Consequently, $S_{SU(4)}$ drops with increasing $U$ and
the residue entropy, in principle, approaches zero.
If $U$ continues to increase and reaches the strong interaction regime, say,
$U/t>8$, $\Delta_c\simeq U/2$ increases while $\Delta_s\simeq J= 4t^2/U$
decreases. Thus the relation $\Delta_c\gg T\gg \Delta_s$ holds.
$T$ is low enough to freeze charge fluctuations but
high enough to disorder AF correlations. Therefore,
increasing $U$ (or equivalently, decreasing $\Delta_s$)
enhances the entropy at a fixed $T$ in the low temperature
regime.

{\it Intermediate temperature regime $T_l< T <T_h$ }
In this temperature regime, $S_{SU(4)}$ monotonically increases with
$U$ at a fixed $T$.
At small $U$ the system is in the weak Mott-insulating
state with small values of $\Delta_s$ and $\Delta_c$, which is
very close to a Fermi liquid state.
The system enters the solid-like strong Mott state at large $U$,
where the physics is mostly local-moment-like.
The increase of entropy with $U$ means that the liquid-like
state is more ordered than the solid-like state, known as the
Pomeranchuk effect.
In this temperature regime, $T$ is high enough compared to $\Delta_s$,
and thus thermal fluctuations suppress magnetic correlations, while
it remains smaller than $\Delta_c$, such that charge fluctuations
are still frozen.
In the strong Mott-insulating state, fermions on each site are nearly
independent of each other, and thus the entropy per site is proportional
to the logarithmic of the spin degeneracy.
In the weak Mott-insulating state, we could think there is still
a reminiscence of Fermi surface, which strongly suppresses
the entropy contribution.
Therefore, the liquid-like state is more ordered than the solid-like state
in the intermediate temperature regime.

On the other hand, the Pomeranchuk effect does not occur in the low
temperature regime ($T\ll \Delta_s$), where thermal
fluctuations are not strong enough to suppress AF correlations.
In this case, the AF correlations between adjacent sites lift spin degeneracy
and lower the entropy in the strong Mott-insulating state.
Consequently, the Pomeranchuk effect is prominent in the
intermediate temperature regime.

{\it High temperature regime $T>T_h$ }
In this case, not only spin fluctuations but also charge fluctuations
are defrozen by thermal fluctuations $T > \Delta_c$.
Charge fluctuations cannot be completely suppressed by $U$, and contribute most to the entropy.
Therefore, increasing $U$ lowers the entropy at a fixed
$T$ in the high temperature regime.

As for the SU(6) case, its behavior of entropy v.s.
$U$ and $T$ is qualitatively identical to the SU(4) case.
Nevertheless, the intermediate temperature regime of the former
is broader than the latter, which means that the Pomeranchuk
effect is more prominent.

\subsection{The Pomeranchuk effect}

\begin{figure}[htb]
\includegraphics[width=0.99\linewidth]{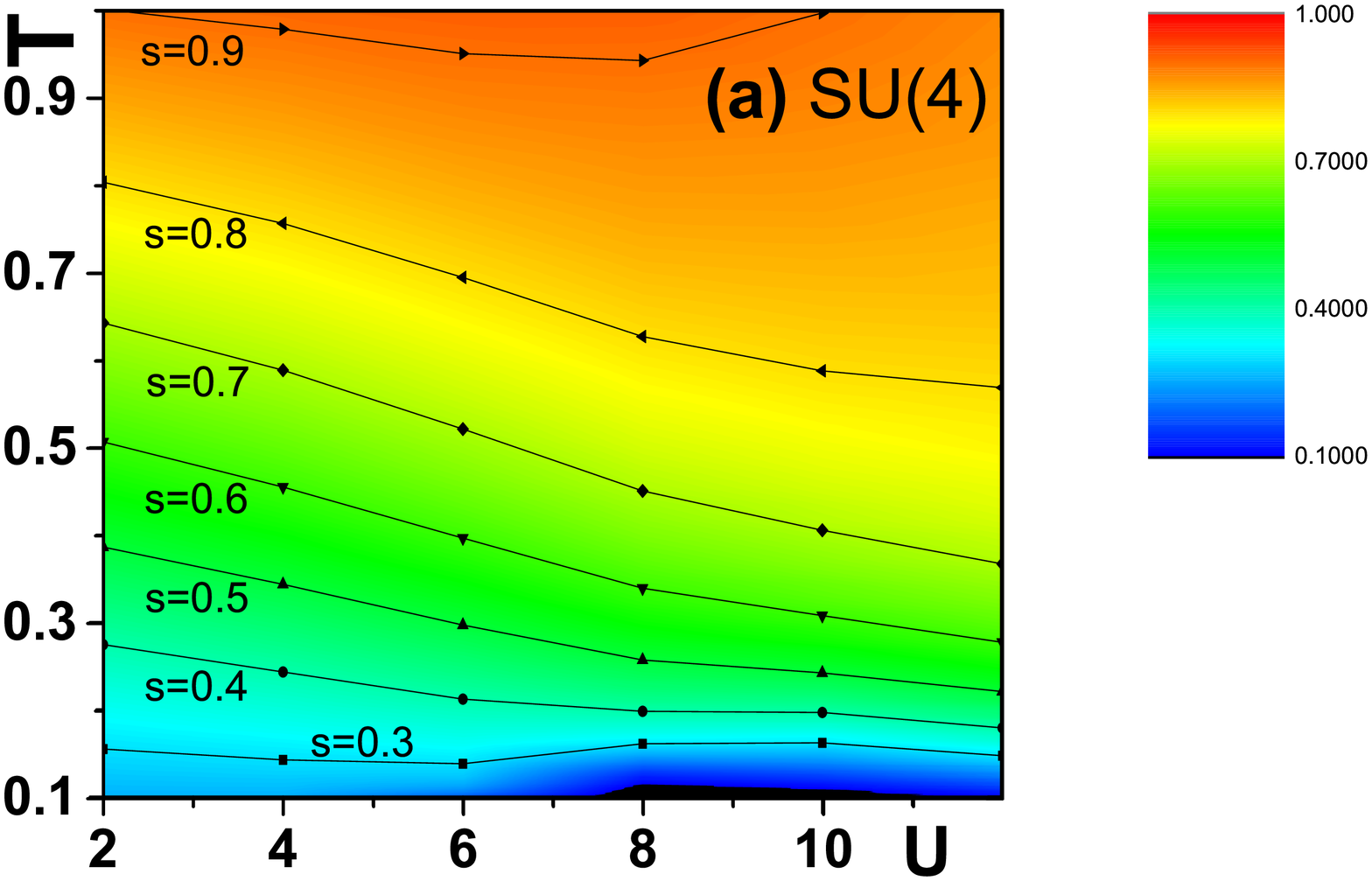}
\includegraphics[width=0.99\linewidth]{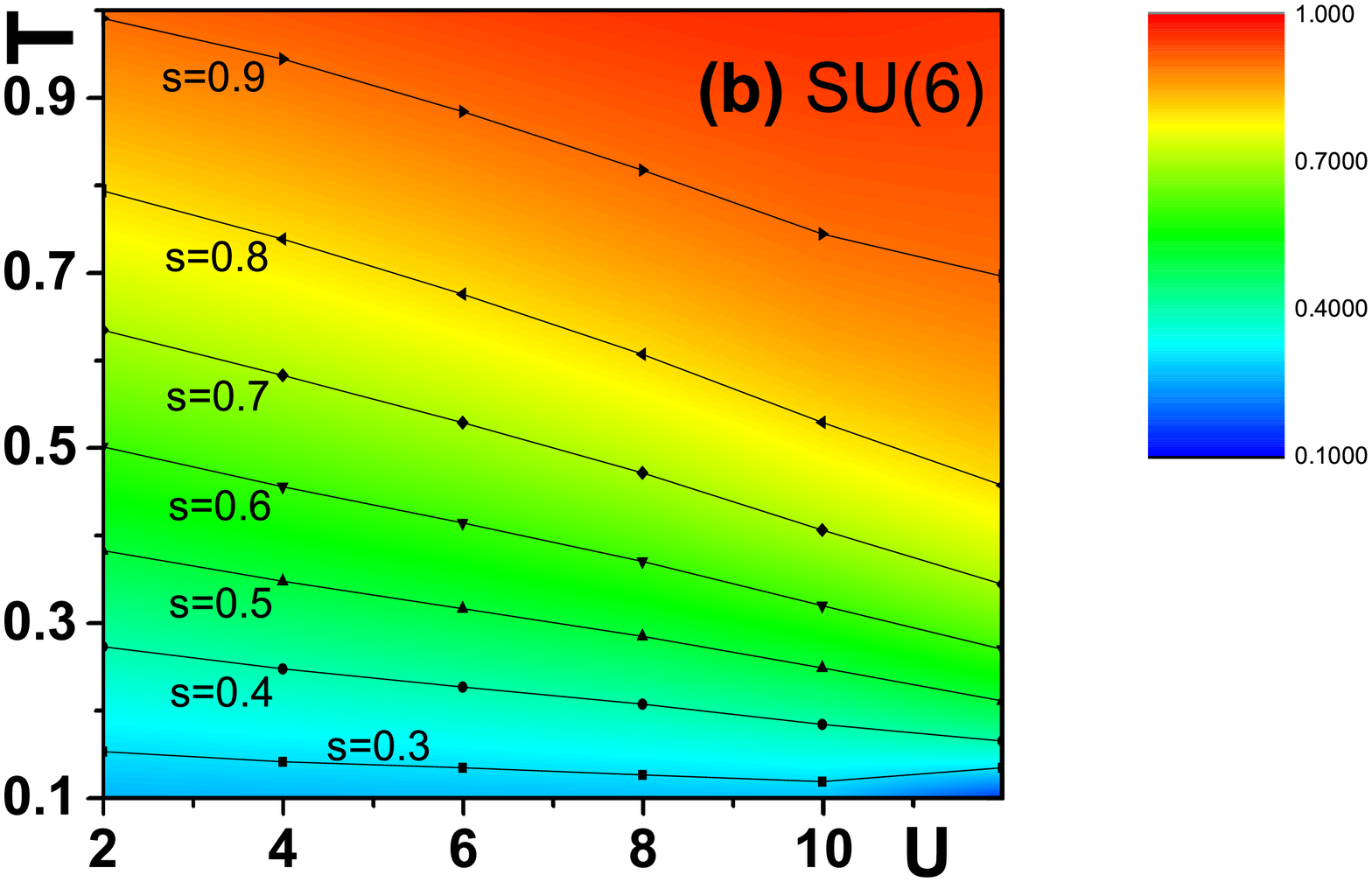}
\caption{The isoentropy curves for the half-filled ($a$)
SU(4) and ($b$) SU(6) Hubbard models on a $10\times 10$ square
lattice. The entropy per particle $S_{SU(2N)}$ is indicated
on each curve.
(b) was published in Ref. [\onlinecite{Cai2013}], which
is reproduced here with a new curve of $S_{SU(6)}=0.3$ added.
}
\label{fig:PC}
\end{figure}

As analyzed in section \ref{sect:entropy}, in the strong Mott-insulating
regime the lattice system has larger entropy capacity than in the weak
Mott-insulating regime, which leads to the Pomeranchuk effect.
This effect was first proposed in the $^3$He system, where,
in the low temperature region, increasing pressure adiabatically
can further cool the system. In low temperature physics,
this effect was employed as an effective cooling method named after
Pomeranchuk.
The similar situation occurs in the Hubbard model.
Nevertheless, at low temperatures where the AF correlations
are important, the spin degeneracy is lifted, which reduces
the entropy in the Mott-insulating state.
In this case, the Pomeranchuk effect does not occur.
The DQMC simulations have been performed for the half-filled SU(2)
Hubbard model in the literature.
Both in 2D and 3D cases, the Pomeranchuk effect is not obvious
in the range of the entropy per particle $S_{SU(2)}$ between $0.1$ to $0.9$, even if
the interaction achieves $U/t\sim 10$ \cite{Dare2007,Paiva2011,Paiva2010}.

In the multi-component SU($2N$) Hubbard model, the situation is different.
Due to the increase of the number of fermion components, the Pomeranchuk
effect is greatly facilitated \cite{Hazzard2010,Cai2013}.
The isoentropy curves v.s. $U$ and $T$ are plotted in Fig. \ref{fig:PC}
(a) and (b) for the SU(4) and SU(6) Hubbard models, respectively.
In both cases, the Pomeranchuk effects are prominent in the
intermediate temperature regime with intermediate interaction
strengths, which emerge at $0.4<S_{SU(4)}<0.8$ for the SU(4) case,
and at $0.3<S_{SU(6)}<0.9$ for the SU(6) case.

The enhancement of the Pomeranchuk effect can be illustrated
by comparing the SU(2) and SU(4) cases.
When deeply inside the Mott-insulating state, in the intermediate
temperature regime, the AF correlations can
be neglected, and the entropy per particle $S_{SU(2N)}$
is dominated by the contribution of spin degeneracy.
Therefore, the entropy capacities can be estimated as $S_{SU(2)}
=\ln 2\approx 0.69$, which is smaller than $S_{SU(4)} =
\ln(C_4^2)/2 \approx 0.89$.
On the other hand, in the intermediate interaction regime, the charge gap
in the SU(4) case is significantly smaller than that in the SU(2) case
for the same interaction $U$.
This means that fermions in the SU(4) Hubbard model can be more easily excited to the upper Hubbard band than those in the SU(2) case,
which also enhances the entropy capacity of fermions in the SU(4) case.

\section{The probability distribution of the onsite
occupation number}
\label{sect:onsite}

\begin{figure}[htb]
\includegraphics[width=0.8\linewidth]{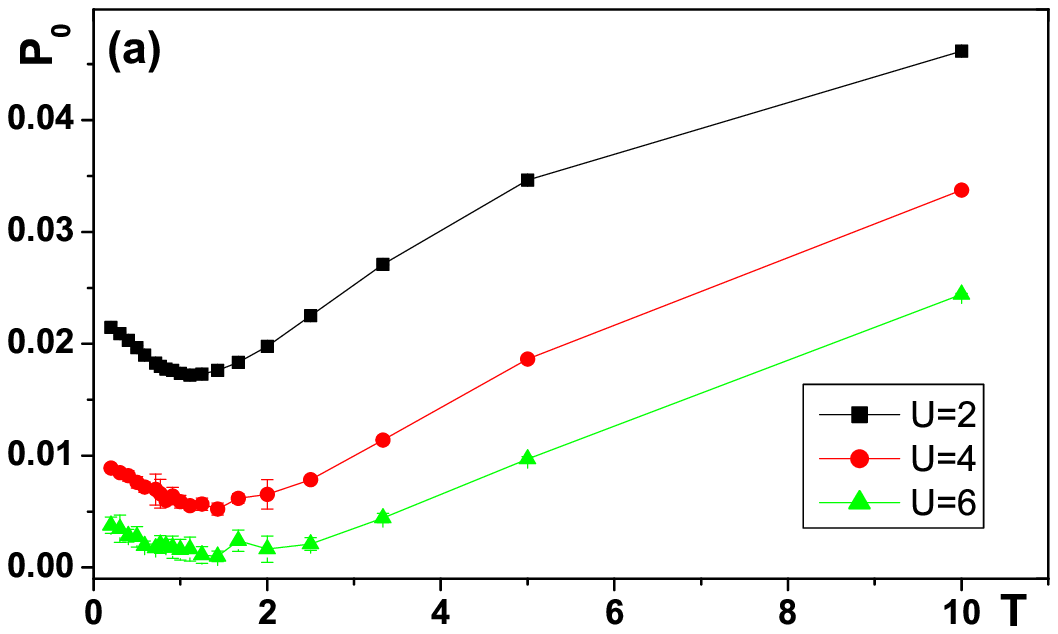}
\includegraphics[width=0.8\linewidth]{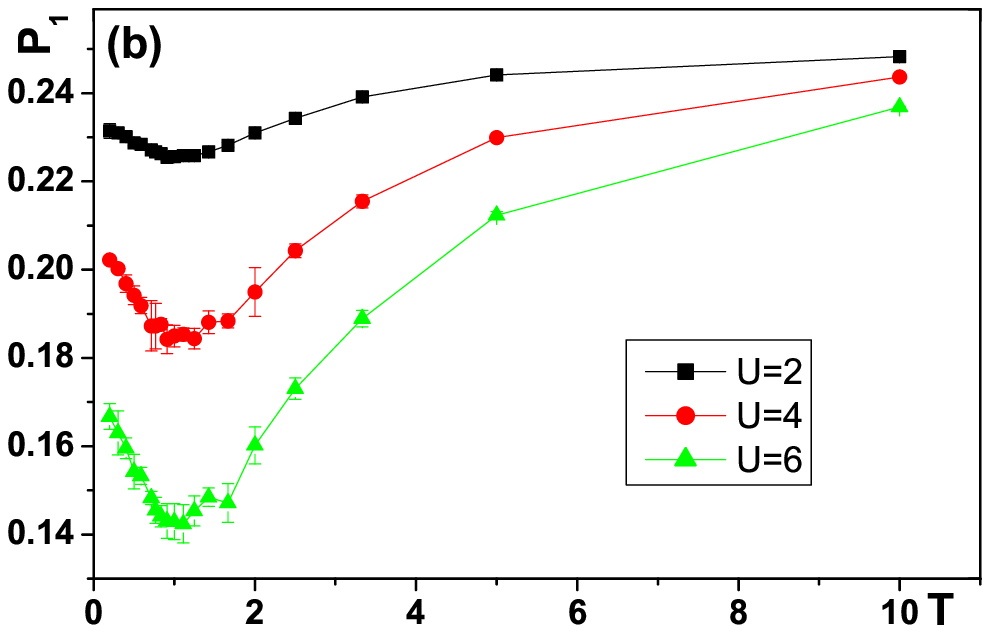}
\includegraphics[width=0.8\linewidth]{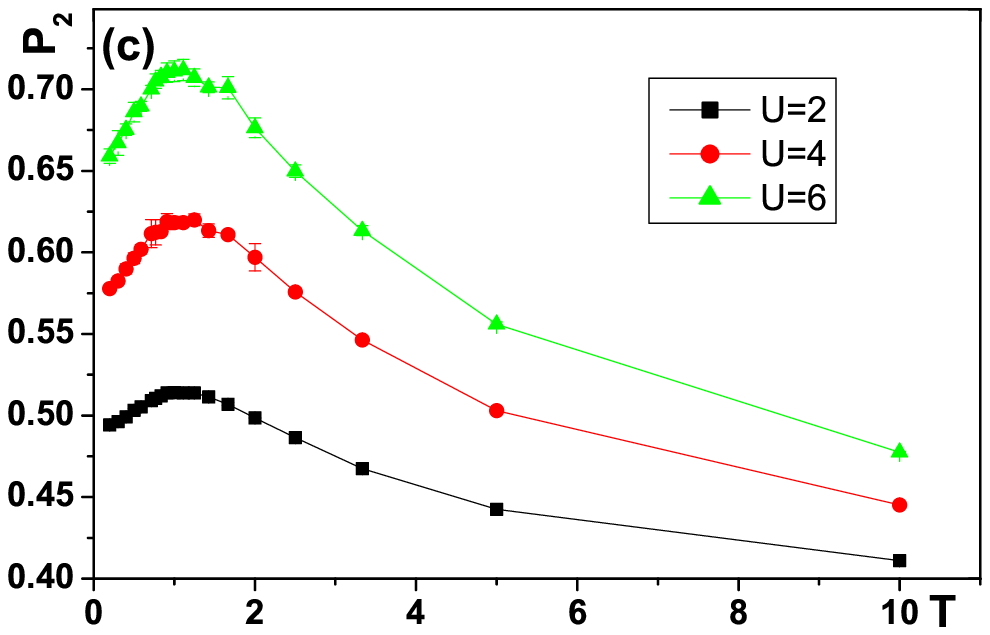}
\caption{The probabilities $P(n)$ for the onsite particle
numbers (a) $P(0)$, (b) $P(1)$, and (c) $P(2)$ v.s.
$T$ and $U$ in the half-filled SU(4) Hubbard model.
Due to the particle-hole symmetry, $P(0)=P(4)$, and
$P(1)=P(3)$, and $2P(0) +2 P(1)+ P(2)=1$.
The lattice size is $L\times L$ with $L=10$. }
\label{fig:momentum}
\end{figure}

To characterize charge fluctuations, we study the probability distribution
of the onsite occupation number.
In the SU(2) case, at half filling the double-occupation number $n_d(i)=\langle
n(i)_\uparrow n(i)_\downarrow\rangle$ is related to the local
moment $\langle m^2_z\rangle=1-2n_d$, which
exhibits a slightly non-monotonic behavior as a function of $T$
for fixed $U$\cite{Paiva2001}.
Also, in cold atom experiments, this quantity can be
measured with {\it in situ} single-site resolution techniques
\cite{Bakr2010,Sherson2010}.
Let us consider the SU(4) case. At half-filling the most probable
configuration of onsite particle number is $n(i)=2$.
At finite $U$, particles are allowed to hop between different
sites, leading to charge fluctuations.
Due to particle-hole symmetry, the probabilities for the occupation
numbers $n$ and $2N-n$ are equal. Thus we
only need to calculate $P(n)$ with $n=0,1$, and $2$, respectively.
They are defined as
\begin{eqnarray}
P(0)&=&\prod_{\alpha=1}^4(1-n_i^\alpha);\nn \\
P(1)&=&\sum_{\alpha=1}^4 n_i^{\alpha}\prod_{\beta\neq \alpha} (1-n_i^{\beta}); \nn \\
P(2)&=&\sum_{\alpha\neq\beta}
n_i^{\alpha}n_i^{\beta}\prod_{\gamma\neq\alpha\,\beta}(1-n_i^{\gamma}).
\end{eqnarray}
Obviously, they satisfy the relation $2P(0)+2P(1)+P(2)=1$.

The simulation results of $P(n)$ ($n=0,1$ and $2$) as a function of $T$
for various $U$ are presented in Fig. \ref{fig:momentum}.
In the weak and intermediate interaction regimes, charge fluctuations
are significant.
The maximal probability of the exact half-filling $P(2)$ only achieves
around 70\% when $U/t=6$.
In contrast, the probabilities of one-particle fluctuation defined by $2P(1)$
fall in the range between $30\%$ and $40\%$.
The probabilities of two-particle fluctuation, $2P(0)$, are typically as low as
a few percent.
Each $P(n)$ ($n=0,1$ and $2$) exhibits non-monotonic behavior
as $T$ increases.
For example, at low temperatures $P_0$ and $P_1$ fall with the increase of
$T$; then after reaching their minima at the temperature scale around
$t$, they grow again.
This indicates that in the temperature regime $T<t$, on-site charge fluctuations are suppressed with
increasing $T$.
This counterintuitive phenomenon reminds us of the Pomeranchuk effect,
that the system tends to localize fermions in the intermediate
temperature regime to maximize the entropy, which mainly comes from
the spin degeneracy. When the temperature further increases,
charge fluctuations are activated and grow with $T$.

\section{The magnetic properties}
\label{sect:mag}

In this section, we study the magnetic properties of the half-filled SU($2N$) ($2N$=4 and 6)
Hubbard model on a square lattice at finite temperatures. including
both the uniform spin susceptibilities and the AF structure factor.

\subsection{The uniform spin susceptibilities}
\label{sect:sus}

\begin{figure}[htb]
\includegraphics[width=0.9\linewidth]{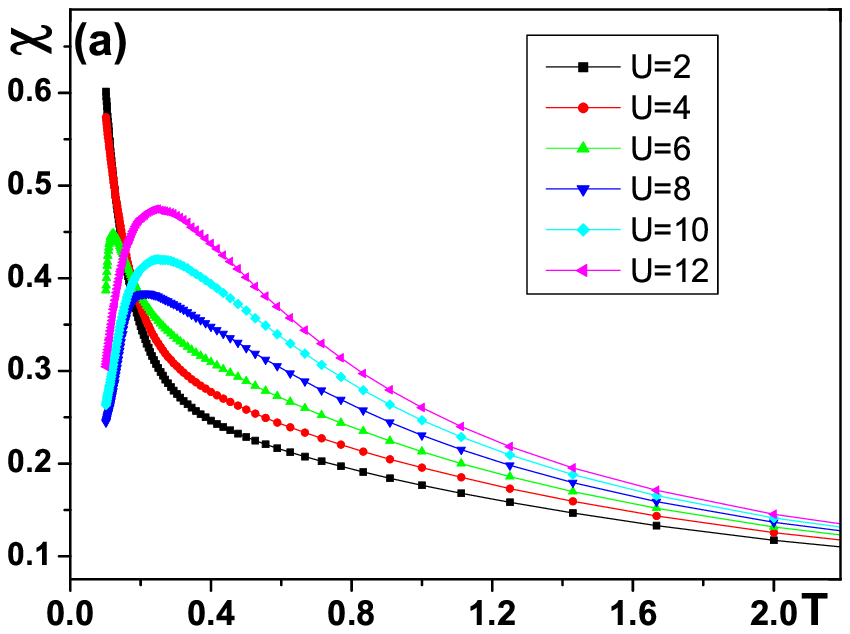}
\includegraphics[width=0.9\linewidth]{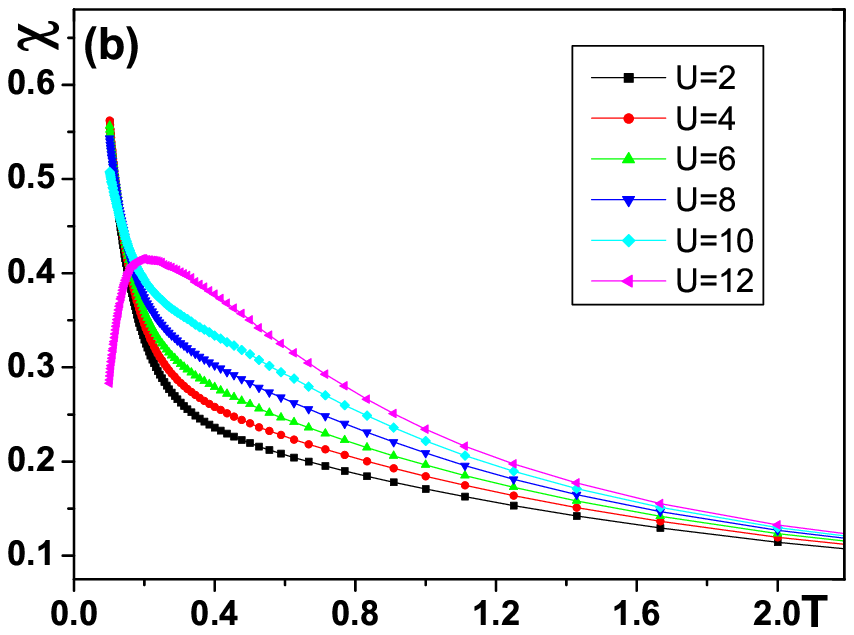}
\caption{The uniform spin susceptibilities v.s. $T$ for various $U$ in the half-filled (a) SU(4) and (b)
SU(6) Hubbard models.
The system size is $L\times L$ with $L=10$.
Error bars are smaller than the data points.
}
\label{fig:chi}
\end{figure}


We now consider the uniform spin susceptibilities $\chi_{SU(2N)}$.
Because the total spin is conserved, $\chi_{SU(2N)}$ can be expressed
as the equal-time correlations through the structure factor
at $\vec q=0$:
\bea
\chi_{su(2N)}(T)=\frac{1}{k_B T} S_{SU(2N)}(\vec q=0).
\eea
where the structure factor $S_{su(2N)}(\vec q)$ is defined
by Eq.(\ref{eq:struc}).

In the 2D half-filled SU(2) Hubbard model \cite{Paiva2011}, or, the
Heisenberg model \cite{Makivic1991}, it is known that at high
temperatures, $\chi_{su(2)}$ behaves as the Curie-Weiss law
which is proportional to $1/T$.
At low temperatures,  $\chi_{su(2)}(T)$ is suppressed due to
the AF correlations, and
therefore it exhibits a peak at a low temperature scale $T_p$.
$T_p$ can be used to roughly characterize the spin energy scale
$T_p \simeq \Delta_s $ (in Heisenberg model $T_p\simeq J$).

The simulation results of the uniform spin susceptibilities for
the SU(4) and SU(6) Hubbard models are presented in Figs. \ref{fig:chi}
(a) and (b), respectively.
Only the weak and intermediate interaction regimes
with $2\le U/t \le 12$ are considered here.
In the SU(4) case, when $U/t<6$, the energy scale of
$\Delta_s$ is very small, and thus the peak location is beyond the
temperature scope $T/t>0.1$ in our simulations.
Compared to the SU(2) case, $\Delta_s$ in the SU(4) case
is significantly weakened.
For example, the peak in the SU(2) Hubbard model is located around
$T_p/t\approx 0.3$ when $U/t=4$ as simulated in Ref. [\onlinecite{Cai2013}].
Nevertheless, the peak locations in the interaction regime $6< U/t <12$ have already
become visible at temperatures $T_p/t>0.1$.
Furthermore, the magnitude of $T_p$ and the peak of $\chi_{SU(4)}$
increases with $U$, which shows the enhancement of AF correlations.
In the SU(6) case, the AF correlations are further weakened
compared to the SU(4) case:
among the curves in Fig. \ref{fig:chi} (b), only the one with $U/t=12$
exhibits a peak visible at temperatures $T_p/t>0.1$.
This indicates the weakening of the spin energy scale
$\Delta_s$ in the intermediate interaction regime with
increasing the number of fermion components.

\subsection{AF structure factors}
\label{sect:afm}

\begin{figure}[htb]
\includegraphics[width=0.9\linewidth]{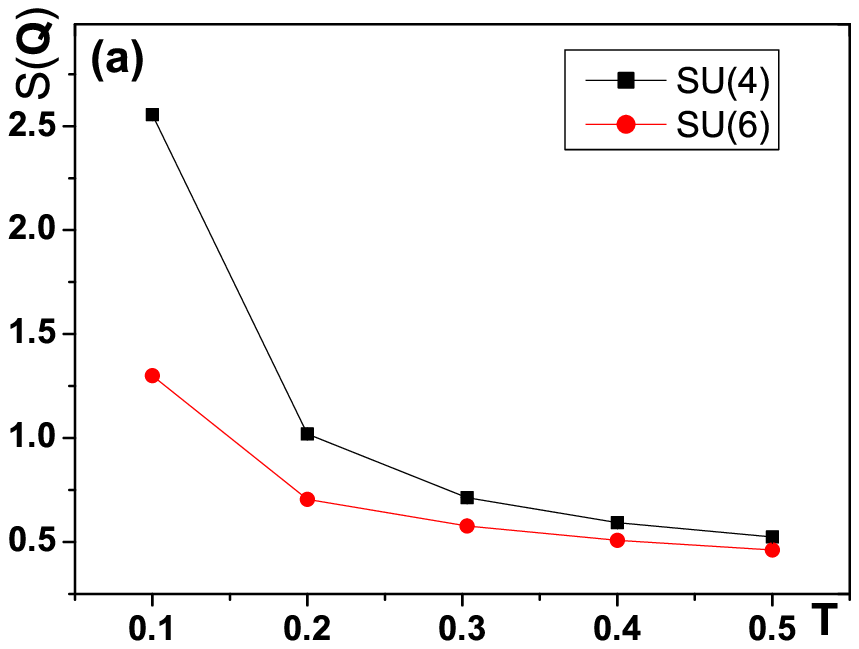}
\includegraphics[width=0.9\linewidth]{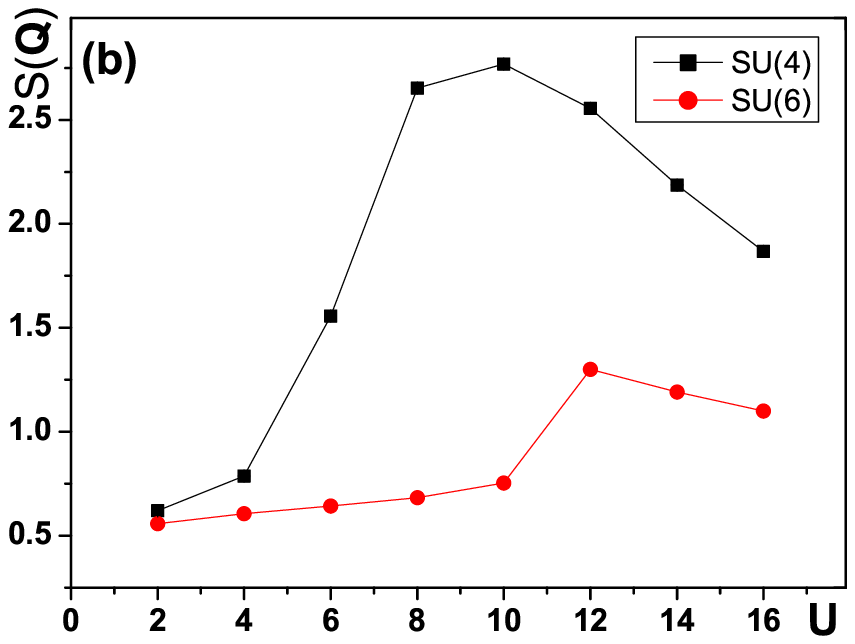}
\caption{The AF structure factors $S_{SU(2N)}(\mathbf{Q})$ with
$\mathbf{Q}=(\pi,\pi)$: (a) as a function of $T$ for fixed $U/t=12$, and
(b) as a function of $U$ for fixed $T/t=0.1$ for the half-filled
SU(4) and SU(6) Hubbard models.
The system size is $L\times L$ with $L=10$.
Error bars are smaller than the data points.}
\label{fig:structure}
\end{figure}

We use the AF structure factor $S_{SU(2N)}(\mathbf{Q})$ defined with staggered
wave vector $\mathbf{Q}=(\pi,\pi)$ to describe the AF correlations.
In Fig. \ref{fig:structure} (a), the curves of the AF structure factor v.s. $T$ are plotted at $U/t=12$ in both
the SU(4) and SU(6) cases.
Both $S_{SU(4)}(\mathbf{Q})$ and $S_{SU(6)}(\mathbf{Q})$ increase monotonically
with the decrease of $T$, which indicates the development of AF correlations.
At a fixed $T$, $S_{SU(4)}(\mathbf{Q})$ is stronger than
$S_{SU(6)}(\mathbf{Q})$, which becomes even more prominent at low temperatures.
This is consistent with the picture that increasing the number
of fermion components suppresses AF correlations.

In Fig. \ref{fig:structure} (b), the dependence of $S_{SU(2N)}(\mathbf{Q})$
on $U$ at a fixed temperature, $T/t=0.1$, is plotted in both SU(4) and
SU(6) cases, where the $U$-dependencies exhibit a non-monotonic behavior.
At small $U$, $S_{SU(4)}(\mathbf{Q})$ and $S_{SU(6)}(\mathbf{Q})$
increase with $U$.
In this weak interacting regime, $\Delta_c\simeq \Delta_s < T=0.1t$,
and thus increasing $U$ enhances the spin energy scale but suppresses
charge fluctuations, which facilitates to build up the AF correlation.
Again, the enhancement of the AF correlation is more prominent in the SU(4)
case than in the SU(6) case.
An interesting feature is that the rates of increase jump at $U/t\approx 4$
in the SU(4) case, and at $U/t\approx 10$ in the SU(6) case.
This may be due to a rapid increase of the charge energy scale $\Delta_c$,
which indicates the crossover from the weak interaction regime
to the intermediate interaction regime.
The AF in the weak interaction regime is due to Fermi surface nesting,
and it evolves to the local moment physics as $U$ enters the intermediate
interaction regime.
At large $U$, $\Delta_s\simeq J=4t^2/U$ decreases with the increase of
$U$.
Thermal fluctuations are described by the parameter $T/J$, and thus
increasing $U$ effectively enhances thermal fluctuations.

On the other hand, the zero temperature
projector QMC results \cite{Wang2014} show that the AF long-range
orderings reach the maxima around $U/t\approx 8$ and $10$
in the SU(4) and SU(6) cases, respectively.
The AF ordering is then suppressed by further increasing $U$, which
is an effect of quantum fluctuations.
In the SU(6) case, the AF long-range order is even completely suppressed
around $U/t\approx 15$ at zero temperature.
This is because when deeply inside the Mott-insulating regime,
the number of superexchange processes between two adjacent sites
increases rapidly with the number of fermion components, which indicates the enhancement of
quantum fluctuations.
Combining both effects of quantum and thermal fluctuations,
the AF correlations are weakened with increasing $U$ in the strong
interaction regime.

\section{Conclusions}
\label{sect:conclusion}
In summary, we have performed a systematic DQMC simulation study of thermodynamic properties of the half-filled SU($2N$) Hubbard
model on a square lattice.
Various thermodynamic behaviors including the entropy-temperature
relation, the isoentropy curve and the probability distribution of the onsite
occupation number have been simulated, which demonstrate the Pomeranchuk effect is facilitated with increasing fermion
components. Based on the charge and spin energy
scales, we have analyzed the thermodynamic properties in weak and strong interaction regimes.
In the weak interaction regime, the physics is characterized by the Fermi surface
nesting, while in the strong interaction regime the physics is mostly in the local
moment picture. Additionally, in our simulations
the uniform spin susceptibilities and the
AF structure factors both exhibit qualitatively different
behaviors in weak and strong interaction regimes.
Theoretical analysis as well as DQMC simulations show that the interaction strength separating weak and strong interaction regimes
increases with the number of fermion components.

\acknowledgements
Z.Z., C.W., and Y.W. acknowledges financial support from the
National Natural Science Foundation of China under Grant No. 11328403, and the Fundamental Research Funds for the Central Universities.
C.W. is supported by the NSF DMR-1410375 and AFOSR FA9550-14-1-0168.
Z.C. acknowledges the support from SFB FoQuS (FWF Project No.F4006-N16)
and the ERC Synergy Grant UQUAM.


\appendix
\section*{Appendix: An exact Hubbard-Stratonovich decomposition for
SU(4) and SU(6) Hubbard interactions}
\label{append}
For the SU(2) case, the HS transformation is usually
performed by using the discrete Ising fields \cite{Hirsch1983,Hirsch1985}.
However, the decomposition in spin channel can not be easily generalized
to the SU($2N$) case due to the increase of spin components.
Instead, we choose an discrete HS decomposition in the density channel
at the price of involving complex numbers as used in Ref.
[\onlinecite{Assaad1998}].
The HS transformation for a half-filled SU($2N$) Hubbard model reads:
\begin{equation}
e^{-\frac{\Delta\tau U}{2}(n_{j}-N)^{2}}=\frac{1}{4}\sum_{l=\pm1,\pm2}
\gamma_{j}(l)e^{i\eta_{j}(l)(n_{j}-N)},
\end{equation}
where two sets of discrete HS fields $\gamma$ and $\eta$ are employed.

However, the HS decomposition with an error of order $(\Delta \tau)^4$ in Ref.[\onlinecite{Assaad1998}] is
not exact.
In Ref.[\onlinecite{Wang2014}], a new HS decomposition was
proposed with a new set of parameters, which is exact
for the SU($4$) and SU($6$) Hubbard interactions.
The Ising fields are defined as follows
\begin{eqnarray}
\nonumber \gamma(\pm1)&=&\frac{-a(3+a^{2})+d}{d};\\
\nonumber \gamma(\pm2)&=&\frac{a(3+a^{2})+d}{d};\\
\nonumber \eta(\pm1)&=&\pm\cos^{-1}
\left\{ \frac{a+2a^{3}+a^{5}+(a^{2}-1)d}{4}\right\};\\
\nonumber \eta(\pm2)&=&\pm\cos^{-1}
\left\{ \frac{a+2a^{3}+a^{5}-(a^{2}-1)d}{4}\right\}
\end{eqnarray}
where $a=e^{-\Delta\tau U/2}$, and $d=\sqrt{8+a^{2}(3+a^{2})^{2}}$.


\end{document}